\documentstyle[axodraw,cite]{aipproc}
\newcommand{\mA}{m_{\scriptscriptstyle A}}
\newcommand{\mB}{m_{\scriptscriptstyle B}}
\newcommand{\LO}{\Lambda^0}
\newcommand{\SO}{\Sigma^0}
\newcommand{\SM}{\Sigma^-}
\newcommand{\SP}{\Sigma^+}
\newcommand{\SC}{\Sigma^\pm}
\newcommand{\XO}{\Xi^0}
\newcommand{\XM}{\Xi^-}
\newcommand{\vev}[1]{\left\langle#1\right\rangle}
\newif\ifpreprint \preprinttrue
\ifpreprint
  
\fi
\begin{document}
\title{%
  Hyperon Beta-Decay Analysis \\
  and the Recent KTeV Data%
  \ifpreprint \thanks{Presented at SPIN 2000---Osaka, Oct.~2000} \fi
}

\author{Philip G. Ratcliffe}

\address{%
  Dipartimento di Scienze CC.FF.MM., \\
  Universit{\`a} degli Studi dell'Insubria---sede di Como \\
  via Valleggio 11, 22100 Como, Italy \\
  and \\
  Istituto Nazionale di Fisica Nucleare---sezione di Milano \\
  via Celoria 15, 20133 Milano, Italy \\[0.5ex]
  \texttt{\upshape pgr@fis.unico.it}
}

\ifpreprint \date{\vspace*{4ex}\noindent\hfil December 2000} \fi

\maketitle
\begin{abstract}
  The analysis of hyperon semi-leptonic decay data is addressed with
  reference to SU(3) breaking and isospin mixing between $\LO$ and $\SO$.
  Various approaches to SU(3) breaking are discussed and compared. The
  phenomenological implications of $\LO{-}\SO$ mixing are not to be
  underestimated: it can induce vector couplings in decays otherwise purely
  axial and may also modify rates. In regard of the KTeV data on
  $\XO\to{\SP}e\bar\nu$, predictions are presented and the impact of present
  a future data on the extraction of $F$ and $D$ is also examined. In
  addition, the implications of the new data for the use of octet baryon beta
  decays in determining $V_{us}$ are considered.
\end{abstract}

\ifpreprint \newpage \fi
\section{Introduction}
Hyperon semi-leptonic decay (HSD) data are the sole present source of
information on the $F$ and $D$ parameters, vital for the analysis of polarised
deep-inelastic scattering experiments. In addition, they may provide access to
the Cabibbo-Kobayashi-Maskawa (CKM) matrix element, $V_{us}$. Recent data for
$\XO\to{\SP}e\bar\nu$ from the KTeV collaboration \cite{Affolder:1999pe} at
Fermilab have added new interest to this area.

Beside the accurately measured neutron $\beta$-decay rate and angular
asymmetries, there is a body of data on the rest of the baryon octet
\cite{Groom:2000in}. In SU(3) such decays are described via two parameters,
$F$ and $D$, relating to strong-interaction effects and two further
parameters, $V_{ud}$ and $V_{us}$, the CKM matrix elements (heavy-flavour
contributions may be neglected). The $F$ and $D$ parameters are important in
connection with in the Ellis-Jaffe sum rule \cite{Ellis:1974kp}; a 15\%
reduction in the ratio $F/D$ from its accepted value ($\sim\!0.6$) would
remove the discrepancy with polarised DIS data and alleviate the ``proton-spin
puzzle'' \cite{Close:1993mv}.

As SU(3) is violated at about the 10\% level, a reliable description of the
breaking is important. A test of any scheme lies in the predictions made for
new decays: such as, the process $\XO\to\SP{e}\bar\nu$, which has now been
measured by the KTeV collaboration \cite{Affolder:1999pe,
Alavi-Harati:1999tq}. In this talk, after outlining the data and an approach
to SU(3) breaking, I shall present a prediction for this decay
\cite{Ratcliffe:1998su} and discuss future developments.
\section{The HSD Data}
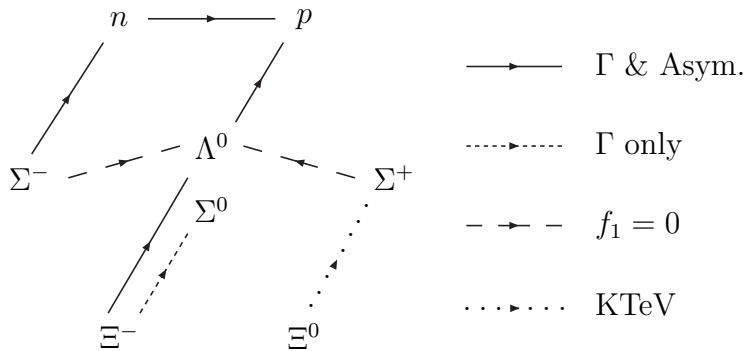
\begin{figure}[hbt]
\centering
\begin{picture}(200,140)(-45,-70)
  \SetWidth{1.0} \SetScale{0.6}
  \ArrowLine    (-113,  15)( -68,  85)      
  \ArrowLine    (  15,  35)(  45,  85)      
  \ArrowLine    ( -65, -85)( -15,   0)      
  \ArrowLine    ( -40, 100)(  40, 100)      
  \DashArrowLine( -90,   0)( -20,  20){15}  
  \DashArrowLine(  90,   0)(  20,  20){15}  
  \DashArrowLine( -45, -85)( -15, -35){3}   
  \put          (  36, -49){.}
  \put          (  39, -44){.}
  \put          (  42, -38){.}
  \ArrowLine    (  77, -55)(  78, -53)
  \put          (  48, -28){.}
  \put          (  51, -22){.}
  \put          (  54, -17){.}
  \put          (  57, -11){.}
  \Text         ( -35,  60)[c]{$n$}
  \Text         (  35,  60)[c]{$p$}
  \Text         ( -69,   0)[c]{$\SM$}
  \Text         (   0,  12)[c]{$\LO$}
  \Text         (   0, -12)[c]{$\SO$}
  \Text         (  69,   0)[c]{$\SP$}
  \Text         ( -35, -60)[c]{$\XM$}
  \Text         (  35, -60)[c]{$\XO$}
  \ArrowLine    ( 160,  70)( 220,  70)
  \DashArrowLine( 160,  20)( 220,  20){3}
  \DashArrowLine( 160, -30)( 220, -30){15}
  \put          (  95, -50){.}
  \put          ( 101, -50){.}
  \put          ( 107, -50){.}
  \ArrowLine    ( 188, -82)( 193, -82)
  \put          ( 119, -50){.}
  \put          ( 125, -50){.}
  \put          ( 131, -50){.}
  \Text         ( 145,  42)[l]{$\Gamma$ \& Asym.}
  \Text         ( 145,  12)[l]{$\Gamma$ only}
  \Text         ( 145, -18)[l]{$f_1=0$}
  \Text         ( 145, -48)[l]{KTeV}
\end{picture}
\caption{%
  The SU(3) scheme of the measured baryon octet $\beta$-decays: the solid
  lines represent decays for which both rates and asymmetry measurements are
  available; the short dash, only rates; the long dash, $f_1=0$ decays; and
  the dotted line, the KTeV measurement.}
\label{SU3scheme}
\end{figure}

A number of hyperon semi-leptonic decays have been measured with varying
degrees of accuracy and depth of information; fig.~\ref{SU3scheme} depicts
the measured baryon octet $\beta$-decays, indicating the type of data
available, while the present world HSD data are collected in
table~\ref{tab:HSDdata}. Note that several of the rates and asymmetries have
now been measured to better than 5\%. Moreover, I should add that a past
discrepancy in the neutron $\beta$-decay data has now been resolved.
\begin{table}[hbt]
  \caption{
    Present world HSD rate and angular-correlation data
    \protect\cite{Groom:2000in}. Numerical values marked $g_1/f_1$ are
    extracted from angular and spin correlations.}
  \label{tab:HSDdata}
  \begin{tabular}
    {c@{$\;\to\;$}ll@{$\;\pm\;$}ll@{$\;\pm\;$}ld@{$\;\pm\;$}ll}
    \multicolumn{2}{c}{Decay}                   &
    \multicolumn{4}{c}{\qquad Rate ($10^6$\,s$^{-1}$)} &
    \multicolumn{2}{c}{\qquad $g_1/f_1$}               &
    $g_1/f_1$
  \\
    \cline{3-6} \raisebox{0.5ex}{\strut}
    $A$   & $B\ell\nu$ &
    \multicolumn{1}{r@{$\;=\;$}}{$\ell$} & {$e^\pm$} &
    \multicolumn{1}{r@{$\;=\;$}}{$\ell$} & {$\mu^-$} &
    \multicolumn{1}{r@{$\;=\;$}}{$\ell$} & {$e^-$} &
    SU(3)
  \\
    \tableline
    $n$   & $p$   & 1.1278 & 0.0024
    \tablenote{Rate given in units of $10^{-3}\,$s$^{-1}$.} &
    \multicolumn{2}{c}{}           &
    1.2670 & 0.0035
    \tablenote{Scale factor 1.9 included in error (PDG practice for discrepant
      data).} & $F+D$
  \\
    $\LO$ & $p$   & 3.161  & 0.058 & 0.60 & 0.13 &
    0.718 & 0.015 & $F+\frac13D$
  \\
    $\SM$ & $n$   & 6.88   & 0.23  & 3.04 & 0.27 &
    $-$0.340 & 0.017 & $F-D$
  \\
    $\SM$ & $\LO$ & 0.387  & 0.018 & \multicolumn{2}{c}{} &
    \multicolumn{2}{c}{} & $-\sqrt{\vphantom{0}}\frac23\,D$
    \tablenote{Absolute expression for $g_1$ given (as $f_1=0$).}
  \\
    $\SP$ & $\LO$ & 0.250  & 0.063 & \multicolumn{2}{c}{} &
    \multicolumn{2}{c}{} & $-\sqrt{\vphantom{0}}\frac23\,D$
    $^{\mathrm{c}}$ 
  \\
    $\XM$ & $\LO$ & 3.35   & 0.37
    \tablenote{Scale factor 2 included in error (as above).} &
    2.1 & 2.1
    \tablenote{Data not used in these fits.} &
    0.25 & 0.05 & $F-\frac13D$
  \\
    $\XM$ & $\SO$ & 0.53   & 0.10  & \multicolumn{2}{c}{} &
    \multicolumn{2}{c}{} & $F+D$
  \\
    \tableline
  \end{tabular}
\end{table}
\section{SU(3) Breaking and Fit Results}
SU(3) breaking is well described using centre-of-mass (CoM), or recoil,
corrections \cite{Donoghue:1987th, Ratcliffe:1990dh, Ratcliffe:1996fk}. One
approach ($A$ here) is to account for the extended nature of the baryon by
applying momentum smearing to its wave function. For the decay
$A{\to}B\ell\nu$, CoM corrections to $g_1$ lead to
\begin{displaymath}
  g_1
  =
  g_1^{\mathrm{SU(3)}}
    \left[ 1 - \frac{\vev{p^2}}{3\mA\mB}
      \left( \frac14 + \frac{3\mB}{8\mA} + \frac{3\mA}{8\mB} \right)
    \right].
\end{displaymath}
A similar approach ($B$) relates the breaking to mass-splitting effects in the
interaction Hamiltonian via first-order perturbation theory
\cite{Ratcliffe:1997ys}. The correction here takes on the following form:
\begin{displaymath}
  g_1 = g_1^{\mathrm{SU(3)}}\,\left[1-\epsilon(\mA+\mB)\strut\right].
\end{displaymath}
Both approaches normalise the corrections to the reference-point correction
for $g_1^{n{\to}p}$ and depend a single new parameter ($\vev{p^2}$ or
$\epsilon$). Corrections to $f_1$ are negligible in $A$ and assumed so in
$B$, in accordance with the Ademollo-Gatto theorem. Any further global
normalisation correction to the $|\Delta{S}{=}1|$ rates is disfavoured; in
\cite{Donoghue:1987th} a calculated value of $\sim\!8\%$, was used, this is
excluded by present-day fits.

Table~\ref{table:results} displays the results of three fits: $S$
(symmetric), $A$ and $B$. Note that the value of $V_{ud}$ (and hence
$V_{us}$, fixed here via CKM unitarity) is determined mainly by the
super-allowed nuclear \emph{ft} values. However, when $V_{ud}$ and $V_{us}$
(with or \emph{without} imposition of unitarity) are extracted from HSD data
\emph{alone}, all parameter values remain essentially unchanged. Thus,
unitarity appears to be well respected.
\begin{table}[hbt]
  \caption{\label{table:results}%
    SU(3) symmetric and breaking fits, including $V_{ud}$
    from nuclear \emph{ft}.}%
  \begin{tabular}{@{\qquad}c@{\qquad}c@{\qquad}c@{\qquad}c@{\qquad}c@{\qquad}}
    Fit & $V_{ud}$  & $F$ & $D$ &
    \raisebox{0.3ex}{$\chi^2\!\!$}/\raisebox{-0.2ex}{DoF} \\
    \tableline
    $S$ & 0.9748\,(4) & 0.466\,(6) & 0.800\,(6) & 2.3 \\
    $A$ & 0.9740\,(4) & 0.460\,(6) & 0.808\,(6) & 0.8 \\
    $B$ & 0.9740\,(4) & 0.459\,(6) & 0.809\,(6) & 0.8 \\
  \end{tabular}
\end{table}
\section{$\LO$ and $\SO$ Mixing}
While at the level of isospin violation itself the effects are obviously
small, their influence in HSD may, in fact, be significant. It has been
pointed out \cite{Karl:1994ie} that isospin violation can induce mixing
between $\LO$ and $\SO$, described via
\begin{eqnarray*}
  \LO &=&\hphantom{-}\cos\phi \Lambda_8 + \sin\phi \Sigma_8, \\
  \SO &=&          - \sin\phi \Lambda_8 + \cos\phi \Sigma_8.
\end{eqnarray*}
The suggested phenomenological mixing angle is $\phi=-0.86^{\circ}$
\cite{Karl:1999te}. Now, consider, \emph{e.g.}, the $\SC\to\LO$ decay: in
exact SU(2) $f_1$ is zero, thus angular or spin correlations vanish here. If,
however, the $\LO$ contains a small admixture of $\SO$, this is no longer
true. While there is no strong signal in the fits for such mixing,
intriguingly, the values returned are around $-0.8^{\circ}\pm0.8^{\circ}$ in both SU(3)
symmetric and broken fits.
\section{A Prediction}
Rates and parameters may now be predicted for any other decay in the octet;
table~\ref{table:predictions} compares the predictions obtained for
$\XO\to\SP{e}\bar\nu$ from the above three fits.
\begin{table}[hbt]
  \caption{The axial coupling, rate and branching ratio ($BR$) for
    $\XO\to\SP{e}\bar\nu$. The errors are those returned by the fitting
    routine.}
  \label{table:predictions}
  \def\qqq{\qquad\quad}
  \begin{tabular}{@{\qqq}c@{\qqq}l@{\qqq}c@{\qqq}c@{\qqq}}
    Fit & \quad$g_1/f_1$ & $\Gamma$ ($10^6\,$s$^{-1}$) & $BR$ ($10^{-4}$) \\
    \tableline
    $S$ & 1.267\,(0)
    \tablenote{Zero error is assigned to $g_1/f_1$ in the symmetric fit as it
      would be essentially that of neutron $\beta$-decay.}
                      & 0.901\,(42) & 2.61\,(09) \\
    $A$ & 1.151\,(27) & 0.796\,(44) & 2.31\,(10) \\
    $B$ & 1.136\,(30) & 0.781\,(46) & 2.26\,(12) \\
    \tableline
  \end{tabular}
\end{table}
Recall that $g_1/f_1=F+D$ for this decay, thus allowing for important cross
checks. The variation between the two SU(3) breaking fits is well within
statistical errors, I therefore combine the two, obtaining the following mean
values:
\begin{displaymath}
  g_1/f_1 = 1.14\pm0.03\pm0.01
  \quad\mbox{and}\quad
  \Gamma  = (0.79\pm0.05\pm0.01) \cdot 10^6\,\mbox{s}^{-1},
\end{displaymath}
where the second error estimates the systematic uncertainty due to the
difference between fits, the corresponding branching ratio is
$BR=(2.29\pm0.12)\cdot10^{-4}$.

Let me now briefly compare with a $1/N_c$ approach
\cite{Flores-Mendieta:1998ii}: the quoted fit there results in a very low
$F/D=0.46$ and for $\XO\to\SP{e}\bar\nu$ predicts
\begin{displaymath}
  g_1/f_1 = 0.91
  \quad\mbox{and}\quad
  \Gamma = 0.68\cdot10^6\,\mbox{s}^{-1}
  \quad\mbox{(fit $B$ of ref.~\cite{Flores-Mendieta:1998ii}),}
\end{displaymath}
Both values are smaller than those presented here, which are in turn smaller
than na{\"\i}ve SU(3). To comprehend the difference between the predictions, note
that the analysis of ref.~\cite{Flores-Mendieta:1998ii} includes
baryon-decuplet non-leptonic decay data, which dominate; and the overall fit
(\emph{i.e.}, $\chi^2$) is poor. However, applied to the HSD data alone, the
results are similar to those reported here \cite{Manohar:0000pc}. These
differences should be distinguishable by an experiment with good statistics,
such as KTeV \cite{Affolder:1999pe}.

The preliminary KTeV results are as follows \cite{Affolder:1999pe,
Alavi-Harati:1999tq}:
\begin{displaymath}
  g_1/f_1 = 1.24^{+0.20}_{-0.17}\pm0.07
  \quad\mbox{and}\quad
  BR=(2.54\pm0.11\pm0.16)\cdot10^{-4},
\end{displaymath}
the errors are expected to be at least halved when the full data set is
analysed. Although not crucial to the $F$ and $D$ determination (owing to low
sensitivity to the breaking scheme), depending upon the final central values,
the KTeV data should have strong impact on the determination of $V_{us}$.
\section{Conclusions}
Before concluding, let me call attention to an all too often overlooked
point: although easier to analyse (no extra corrections are necessary), the
present data for angular correlations alone show \emph{no} evidence of SU(3)
breaking. Furthermore, compared to the full data set, they lack statistical
power. \emph{Only full analyses can display the true picture}
\cite{Ratcliffe:1996fk}. Here I would comment that the observed consistency
with unitarity (\emph{i.e.}, between $V_{ud}$ and $V_{us}$) and the fact that
there is no disagreement between \emph{complete} analyses (with the possible
exception of \cite{Flores-Mendieta:1998ii}) suggest that the PDG
\cite{Groom:2000in} might reconsider this sector as a contending source for
estimating $V_{us}$. Indeed, unitarity appears better satisfied here than in
the $K_{\ell3}$ data.

A complete comprehension of SU(3) breaking is still wanting: witness the
octet-decuplet discrepancy. Moreover, the system is as yet not truly
over-constrained; in this context, I might also mention another decay (already
measured but not accurately) for which large corrections are expected:
namely, $\XM\to\SO{e}\bar\nu$. There too, $g_1/f_1=F+D$, permitting additional
sensitive cross checks, especially in combination with the KTeV results.

Concluding then, I would stress that, while the data do manifest significant
departures from SU(3), and there is even modest evidence for SU(2) mixing, the
mass-splitting driven schemes discussed here provide an adequate description.
That said, there remains much to be understood, \emph{e.g.}, the long-standing
question of second-class currents. Thus, any new \emph{precise} data are more
than welcome and the contribution of the KTeV collaboration will be
invaluable.

\end{document}